\pdfoutput=1
\documentclass{mem}
\usepackage{natbib}
\usepackage{txfonts}
\usepackage{balance}
\usepackage{graphicx}
\usepackage[a4paper]{hyperref}

\begin{document}
\title{A note on magnetized coronae}
\author{R. \,Belmont\inst{1}  \and M. Tagger\inst{2} }
\institute{Centre d'Etude Spatiale des Rayonnements, 9 rue du Colonel Roche, BP44346,  31028 Toulouse Cedex 4, France, \email{belmont@cesr.fr} \and CEA Service d'Astrophysique, UMR "AstroParticules et Cosmologie", Orme des Merisiers, 91191 Gif-sur-Yvette, France
}
\authorrunning{Belmont}
\titlerunning{Magnetized Coronae}
\abstract{X-ray binaries and AGN show observational evidence for magnetized hot plasmas. Despite years of data, very little is known on these {\it coronae} especially on the mechanisms responsible for their heating, and most models simply assume their existence. However, understanding its properties has now become a key issue of the AGN and microquasars modelling. 
Here we consider the effect of a strong vertical magnetic field on the corona AGN and X-ray binaries and show that  its modeling (structure, heating) must be reconsidered. As a first step, we present one mechanism that could extract energy from the accretion disks and deposits it in the coronae: the {\it magnetic pumping}.
\keywords{}}
\maketitle{}

\section{Introduction}
In addition to a blackbody component, spectra from X-ray binaries in low/hard state, or spectra of AGN exhibit a power-law high energy tail produced by the comptonization of soft photons from the disk by a gas of hot electrons ($k_BT \ge 100$ keV) localized in the inner part of the accreting system. When this {\it corona} dominates the emission, almost the whole accretion power of the disk is transferred to the corona. Several geometries have been suggested, as an ADAF-like medium \citep{E97}, a slab corona lying on both side of the accretion disk \citep{HM93}, a patchy corona with many small active regions \citep{G79}, or more recently, a hot spherical ADAF with a thin and cold accretion disk in the innermost region \citep{Liu06}. 

Here, we point out that the existing modelling of coronae is hard to reconcile with a possible large scale structured magnetic field. It is admitted that accreting systems are magnetized but the strength and geometry of the field are uncertain. On this respect, the jets observed in microquasars and AGN provide crucial constrains. Both polarisation measurements in AGN \citep{P99} and MHD models of jets -that reproduce best the observed acceleration and collimation of matter- point out a strong ($\beta<1$), large scale magnetic field anchored in the disk \citep{F97}.

If confirmed, such a field entirely changes the properties of coronae. For instance, the matter of ADAF-like models is supposed to fall almost in free fall below some transition radius. Such a free fall motion is not compatible with a large scale poloidal magnetic field, since a constant advection of the magnetic flux would soon build a strong field in the inner regions whose pressure would prevent any further advection. 
Also, a heating by the reconnection of field loops emerging from a turbulent disk into an active solar-like corona is difficult to reconcile with a strong vertical field that, unlike in the weakly magnetized solar photosphere ($\beta \approx 1$), would prevent the field reversals required for reconnection. 

Here we address the question of the origin of the high temperature of coronae embedded in a large scale, partly poloidal, strong magnetic field. We show that such a field actually links the disk (where the power is) and the corona. We illustrate this link by presenting a mechanism that uses the field lines to extract power from the disk to the corona. 

\section{Magnetic heating}
Although not compatible with most models of corona, a large scale magnetic field anchored in the disk and extending in an overlying corona must play a crucial role in the energy transfer. With such a field, a strong MHD instability can develop and produce a spiral wave within the disk \citep[the Accretion-Ejection Instability,][]{TP99}. Because of the vertical field lines, this disk perturbation is combined with Alfv\'en and compressible waves that propagate in the corona and carry energy. Many processes can damp these waves and heat the corona; most of them rely on resonance mechanisms. Depending on the exact geometry and physical assumptions on the system, different waves and mechanisms are dominant.

We have investigated one specific mechanism that results from the resonance between two frequencies.
\begin{enumerate}
\item Mostly because of gravity, coronal particles oscillate along the vertical field lines with a frequency close to the Keplerian one.
\item The exciting spiral wave rotates typically with a fraction of the Keplerian frequency.
\end{enumerate}

Depending on the exact conditions, this {\it magnetic pumping} can efficiently extract energy from the disk pattern, and transfer it to coronal particles. \\ 
The situation in microquasars is unfavorable. Electrons oscillate too fast and only ions can be accelerated directly, the instability amplitude is weak ($\delta B/B \approx$ 0.1 ) and the Compton cooling is drastically efficient. As a result, magnetic pumping in microquasars can only extract a small fraction of the accretion power (typically $f<0.01$) and is insufficient to explain the coronal temperature \citep{B05}. \\
The central black hole of our Galaxy also has an accretion disk, a probable vertical field, and exhibits emission from a gas of very hot electrons during flares (see A. Goldwurm?s talk). Conditions for the magnetic pumping seem more favorable there. The wave is stronger ($\delta B/B/\approx 1$), relativistic effects decrease the oscillation frequency of electrons, enabling a direct heating, the fraction of the accretion power needed to be transferred to the hot gas is smaller, radiative cooling is less efficient ($\tau_{\rm cool} \approx \tau_{\rm dyn}$).

\section{Conclusion}

There will be no direct observational evidence about the strength and structure of the magnetic field with any of the planned missions. However, indirect observations with Simbol-X will give new constrains. Better observations of the hard energy tail will give access to a better understanding of the link between the corona and the accretion disk \citep[e.g. phase transitions, Quasi-Periodic Oscillations, long term evolution of the magnetic flux threading the disk][]{T04}. Simbol-X will allow to better characterize the high energy spectrum of the electrons at the Galactic center during flares at a closer distance from the central black hole, in particular, the time evolution of the spectral slope.

\bibliographystyle{aa}

\begin{thebibliography}{}
\bibitem[Belmont \& Tagger(2005)]{B05} Belmont, R., \& Tagger, M.\ 2005, CJAAS, 5, 43
\bibitem[Esin et al.(1997)]{E97} Esin, A.~A., et al.\ 1997, \apj, 489, 865
\bibitem[Ferreira(1997)]{F97} Ferreira, J.\ 1997, \aap, 319, 340
\bibitem[Galeev et al.(1979)]{G79} Galeev, A.~A., et al.\ 1979, \apj, 229, 318 
\bibitem[Haardt \& Maraschi(1993)]{HM93} Haardt, F., \& Maraschi, L.\ 1993, \apj, 413, 507
\bibitem[Liu et al.(2006)]{Liu06} Liu, B.~F., et al.\ 2006, \aap, 454, L9 
\bibitem[Perlman et al.(1999)]{P99} Perlman, E.~S., et al.\ 1999, \aj, 117, 2185
\bibitem[Tagger \& Pellat(1999)]{TP99} Tagger, M., \& Pellat, R.\ 1999, \aap, 349, 1003
\bibitem[Tagger et al.(2004)]{T04} Tagger, M., et al.\ 2004, \apj, 607, 410
\end{thebibliography}

\end{document}